# Simple THz phase retarder based on Mach-Zehnder interferometer for polarization control

Hiroki Ueda[1,2,*], Alexej Pashkin[2], Ece Uykur[2], Manfred Helm[2,3], and Stephan Winnerl[2]

[1]*Center for Photon Science, Forschungsstrasse 111, Paul Scherrer Institute, 5232 Villigen-PSI, Switzerland.*

[2]*Institute of Ion Beam Physics and Materials Research, Helmholtz-Zentrum Dresden-Rossendorf, Dresden 01328, Germany.*

[3]*Institut für Angewandte Physik, Technische Universität Dresden, Dresden 01062, Germany.*

[*]Correspondence authors: hiroki.ueda@psi.ch

## Abstract

On-demand polarization control of electromagnetic waves is the fundamental element of modern optics. Its interest has recently been expanded in the terahertz (THz) range for coherent excitation of collective quasiparticles in matters, triggering a wide variety of non-trivial intriguing physics, e.g., anharmonicity, nonlinear coupling, and metastability. Wavelength tunability in THz polarization control is fundamentally important for the resonant excitation of collective modes. Here, we propose and demonstrate a simple and convenient THz phase retarder based on the Mach-Zehnder interferometer to obtain circular polarization. The efficiency of THz polarization conversion is demonstrated by the achieved high polarization degree of more than 99.9% and a large transmission of ~76%. The simple and compact setup allows us to adapt the phase retarder to existing setups readily and will contribute to further exploration of ultrafast science, e.g., chiral phononics.

## Main text

## Introduction

Coherent excitation of collective quasiparticles using light is an intriguing direct approach for investigating condensed matter physics and tailoring emergent phenomena at ultrafast timescales. Abundant types of collective excitations in matters lie in the terahertz (THz) regime, e.g., molecular vibrations, phonons, magnons, and plasmons, which motivated recent technological development in laser science. Intense terahertz pulses allow us to resonantly create a large amplitude modulation in quasiparticle dynamics, which may result in the induction of a thermally inaccessible metastable state [1-3], stabilizing an almost energetically degenerate excited state [4-6], or nonlinear coupling with another quasiparticle excitation, e.g., exemplified by magnon-phonon coupling [7, 8] or nonlinear phononics [9, 10].

Very recently, strong demand for intense circularly polarized THz pulses has appeared due to the rapidly growing field of chiral phononics [11, 12]. Resonantly driving degenerate phonon modes with a circularly polarized THz radiation creates atomic rotations and an associated effective magnetic field via the ultrafast Barnett effect [13-15], which may require coupling with an electronic subsystem or quantum effects for enhancement in the induced magnetic-field strength [16-21]. The most convenient way to achieve the phase retardation of $\lambda/4$ that converts a linear polarization from a THz laser source, e.g., a free-electron laser [22] or a laser-driven optical rectification [23] and difference frequency generation [24], into circular polarization may be the use of a birefringent crystal, which is typically $x$-cut quartz, though it does not convert a broadband THz pulse into circular as it achieves a perfect circular polarization only at a single wavelength in the terahertz range for a given thickness [14]. Several mechanisms of achromatic THz waveplates have been suggested, e.g., silicon gratings [25], stacked parallel metal plates [26], dielectric artificial birefringence grating filled with polymer dispersed liquid crystal [27], magnetic tuning of molecular orientation in liquid crystals [28], metamaterial [29], total internal reflections in a Fresnel's rhomb [30], and stacked waveplates with appropriate thicknesses and orientations [31], though they are typically bulky, complicated in fabrication and alignment, or with low transmission.

Here, we demonstrate a simple wavelength-variable phase retarder based on the Mach-Zehnder interferometer, i.e., using delay lines for two split identical beams. This setup allows us to convert the linear polarization of a THz pulse to circular with wavelength tunability and high transmission (~76%). Compared to the similar method previously reported using two THz pulses with orthogonal linear polarizations generated by optical rectification of signal and idler beams from optical parametric amplification to combine with an appropriate phase shift [32], a significant advantage is the simplicity and convenience. The performance of the setup is manifested by the achieved large circular polarization degree (> 99.9%).

## Experimental and results

We built up the phase retarder and tested it for two different wavelengths of narrowband THz pulses from the free-electron laser FELBE [Free-Electron Lasers at the ELBE (Electron Linear accelerator with high Brilliance and Low Emittance)] in Helmholtz-Zentrum Dresden Rossendorf, Germany. Figure 1(a) shows a schematic picture of the phase retarder. A *p*-polarized FEL beam is split into two beams with orthogonal ±45° polarizations for transmission and reflection by a free-standing wire grid polarizer consisting of tungsten wires with 25 μm thickness and 10 μm spacing (GS57205, Specac Ltd.), and the beams travel along the different interferometer arms. Due to the difference in the arm lengths ($d_1$ and $d_2$), there is a relative phase shift acquired before recombining the beams at another free-standing wire grid polarizer with the same spec as the first one. When the relative phase shift is π/2 or 3π/2, the obtained polarization is circular, i.e., C+ or C−. The interferometer arms must be almost perfectly balanced, e.g., $d_1 \approx d_2$, for ultrashort laser pulses in contrast to a monochromatic continuous THz source. Otherwise, pulses from the interferometer arms would not interfere. A perfect polarization state is achieved only at a single wavelength but practically remains clean with a wavelength-dependent circular polarization degree $|S_3|$ [C+ ($S_3 = +1$) or C− ($S_3 = -1$)] of >99.5% for the narrowband FELBE beam (~2%) within ±3σ from the optimized central frequency, as shown in Fig. 1(b).

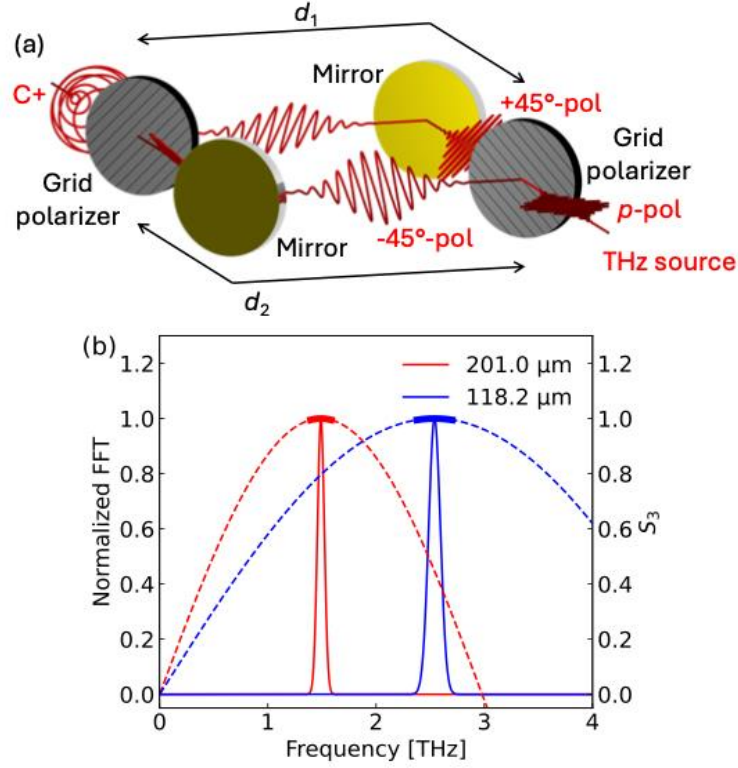

Fig. 1 | (a) A schematic view of the phase retarder based on the Mach-Zehnder interferometer. (b) Normalized spectra of the THz pulse from FELBE (solid curves) and calculated circular polarization degree $S_3$ (dashed curves) for two wavelengths [201 μm (red) and 118.2 μm (blue)] once $S_3$ is optimized for the central frequency. Short thick curves represent $S_3$ spectra for the frequency range within $\pm 3\sigma$ from the central frequency.

We characterized the setup by measuring the transmission intensity as a function of the polarization analyzer angle ($\varphi$) right after the setup. Figures 2(a) and 2(b) show polar plots of the measured intensity after the analyzer for two wavelengths, 118.2 μm (2.54 THz) and 201 μm (1.49 THz), respectively. The experimental data $I(\varphi)$ fits well with

$$I(\varphi) = A_p \cos^2(\varphi - \varphi_0) + A_s \sin^2(\varphi - \varphi_0), \qquad (1)$$

where $\varphi$ represents the analyzer rotation angle, $\varphi_0$ is an offset in the origin of $\varphi$ due to experimental errors, and $A_{p(s)}$ denotes the THz transmission power amplitude along the horizontal (vertical) direction. The fitted parameters allow us to obtain the linear polarization ($\sqrt{S_1^2 + S_2^2}$) degree and circular polarization degree ($|S_3|$) using the so-called Stokes parameters $\mathbf{S} = (S_1, S_2, S_3)$ based on the following equations,

$$\sqrt{S_1^2 + S_2^2} = \left|\cos\left(2\tan^{-1}\frac{A_s}{A_p}\right)\right| \text{ and} \qquad (2)$$

$$|S_3| = \left|\sin\left(2\tan^{-1}\frac{A_s}{A_p}\right)\right|. \qquad (3)$$

Here, the respective Stokes parameters represent the degree of vertical or horizontal linear polarization [$s$ ($S_1 = +1$) or $p$ ($S_1 = -1$)], ±45° linear polarization [+45° ($S_2 = +1$) or −45° ($S_2 = -1$)], and circular polarization, while keeping $|\mathbf{S}| = 1$. Table 1 tabulates the linear and circular polarization degrees for respective nominal polarization states at the two wavelengths. For figures of merit, the achieved circular polarization degrees are larger than 99.9% for all the settings we tested, which demonstrates the high efficiency of our simple setup based on the Mach-Zender interferometer as a phase retarder. Note that the typically obtained linear polarization degree of nominal linear polarization is lower than the circular polarization degree of nominal circular polarization since the phase shift was not as thoroughly optimized, but it is still more than 90%. The transmission of the setup is more than ~76%. Ideally, the Mach-Zehnder interferometer can reach 100% transmission. The lower experimental transmission is probably due to intrinsic losses in the wire grid polarizers and imperfections in their orientation or alignment, resulting in a small portion of the beam leakage in the other direction of the interferometer arms.

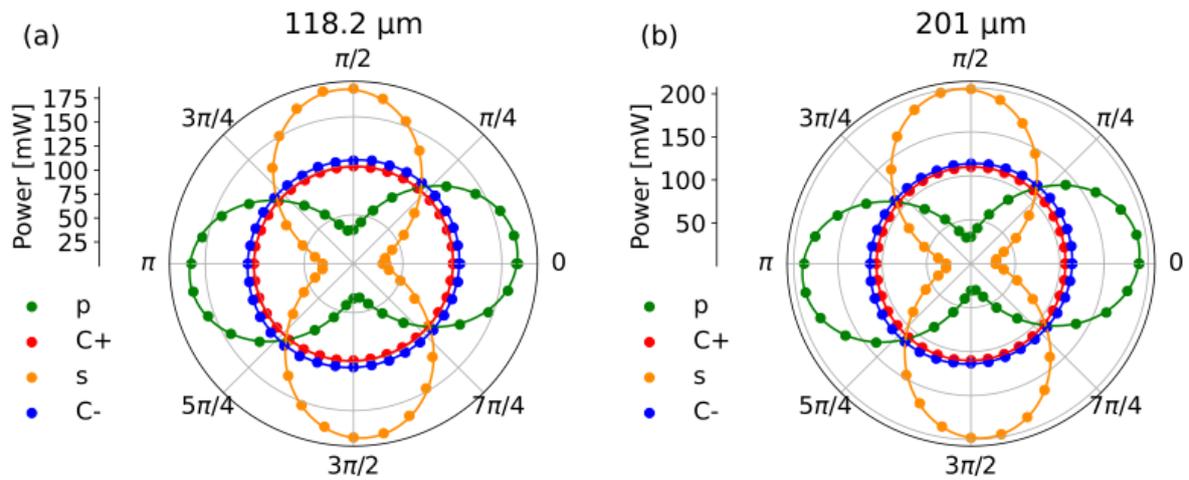

Fig. 2 | Analyzer rotation angle dependence of the transmitted THz power for wavelengths of (a) 118.2 μm and (b) 201 μm for different nominal settings, i.e., linear polarizations [$s$ (orange) and $p$ (green)] and circular polarizations [C+ (red) and C− (blue)]. Filled circles represent experimental data, and curves are fits with Eq. 1 (see main text).

Table 1 | Obtained polarization degrees for respective nominal polarization states and wavelengths.

| Wavelength | Polarization | $\sqrt{S_1^2 + S_2^2}$ | $|S_3|$ |
|---|---|---|---|
| 118.2 μm | $p$ | 91.8% | 39.7% |
|        | C+  | 2.7%  | 100.0% |
|        | s   | 94.7% | 32.2%  |
|        | C−  | 3.7%  | 99.9%  |
| 201 μm | p   | 95.3% | 30.3%  |
|        | C+  | 3.6%  | 99.9%  |
|        | s   | 96.5% | 26.4%  |
|        | C−  | 3.5%  | 99.9%  |

## Conclusions

We have demonstrated the high efficiency of our phase retarder based on the Mach-Zehnder interferometer to convert the polarization of a THz pulse from linear to circular. Even though perfect polarization is achieved only at a single wavelength, the wavelength-dependent polarization degree variation is not relevant for narrowband THz pulse like the one at the FELBE. Compared to the conventional method to control THz polarization by using a quartz waveplate, our method has significant advantages of (1) variable wavelength and (2) wider operation frequency range. A quartz waveplate suffers significant absorption in the mid-infrared range above ~7.5 THz due to optical phonons. However, our setup can operate at any frequency unless the precesion of the adjustment in the split delay lines limits. Our simple phase retarder allows us to explore material dynamics triggered or monitored by polarization-controllable THz pulses.

## Acknowledgments

This research was carried out at ELBE at the Helmholtz-Zentrum Dresden-Rossendorf e. V., a member of the Helmholtz Association. We would like to thank A. Wagner and the FELBE team for their dedicated support.

## Competing interests

The authors declare no competing interests.

## Data availability

Experimental data are accessible from the PSI Public Data Repository [33].